\documentclass[aps,prl,amsmath,twocolumn,PRL,showpacs]{revtex4}

\usepackage{epsfig}

\begin{document}

\title{
Geometrical effects on the optical properties of quantum dots doped
with a single magnetic atom. }

\author{Y.L\'eger}
\email{yleger@spectro.ujf-grenoble.fr}
\author{L.Besombes}
\author{L.Maingault}
\author{D.Ferrand}
\author{H.Mariette}
\affiliation{CEA-CNRS group "Nanophysique et Semiconducteurs",
Laboratoire de Spectrom\'etrie Physique, CNRS and Universit\'e
Joseph Fourier-Grenoble 1, BP 87, F-38402 St Martin d'H\`eres,
France}

\begin{abstract}
The emission spectra of individual self-assembled quantum dots
containing a single magnetic Mn atom differ strongly from dot to
dot. The differences are explained by the influence of the system
geometry, specifically the in-plane asymmetry
 of the quantum dot and the position of the Mn atom. Depending on both
these parameters, one has different characteristic emission features
which either reveal or hide the spin state of the magnetic atom. The
observed behavior in both zero field and under magnetic field can be
explained quantitatively by the interplay between the
exciton-manganese exchange interaction (dependent on the Mn
position) and the anisotropic part of the electron-hole exchange
interaction (related to the asymmetry of the quantum dot).

\end{abstract}


\maketitle

Precise control of electronic spins in semiconductors should lead to
development of novel electronic systems based on the carriers' spin
degree of freedom. These so-called spintronics devices, combining
manipulation of charges and manipulation of spins, could complement
or replace existing electronic systems, yielding new performances
\cite{kane98,loss98}. Magnetic semiconductor quantum dots (QDs),
where excitons (electron-hole pairs) can interact strongly with the
magnetic atoms, hold particular promise as building blocks for such
spin-based systems. However, the geometric factors that become more
and more important with decreasing QD size (because of the quantum
confinement) need to be considered with great care \cite{bayer99}.
For instance, any in-plane asymmetry can introduce very strong
effects in the case of small dots (energy shift of the quantum
levels, induced linear polarization) as revealed by single dot
optical spectroscopy \cite{kul99,bes00}. Thus it is crucial, with
nanometric scale magnetic objects, to understand and to control all
the geometric parameters which characterize the system.

In the case of a quantum dot incorporating a single magnetic atom
(spin S) and a single confined exciton, the exchange interaction
between the exciton and the magnetic atom acts as an effective
magnetic field, so that the atom's spin levels are split even in the
absence of any applied magnetic field \cite{bes04}. A set of $2S+1$
dicrete emission lines can be resolved, providing a direct view of
the atom's spin state at the instant when the exciton annihilates.
However, there has been no experimental study of how a non-ideal dot
geometry, especially departures from circular or square symmetry for
self-assembled QDs grown by Molecular Beam Epitaxy (MBE), affects
optical monitoring of the magnetic spin state.

This letter concerns quantum dots containing a single manganese
magnetic atom. We will show that the electron-hole-Mn system, is
very sensitive to the geometry, specifically the asymmetry of the QD
and the position of the Mn atom in the dot. We report experimental
results showing the three different types of spectra possible.
Control of both the exciton-manganese (X-Mn) exchange interaction
(determined by the position of the Mn) and the anisotropic
electron-hole (e-h) exchange interaction (related to the shape of
the dot) appears as a key condition for detection and manipulation
of the spin state of the isolated magnetic atom. A strong
interaction between the exciton and the Mn splits the six  Mn spin
components, but a strong anisotropy of the dot perturbs the spectrum
pattern and can hide the information on the Mn spin state.

Single Mn atoms are introduced into CdTe/ZnTe QDs during their
growth by MBE, adjusting the density of Mn atoms to be roughly equal
to the density of dots \cite{tin03}. The samples provide symmetric
(disc-like) dots and asymmetric (ellipsoidal) dots, containing Mn
atoms at various positions. The QD emission (at $\simeq2$eV) is
studied in magnetic fields (B=0 to 11T) by optical
micro-spectroscopy in Faraday configuration under non resonant laser
excitation.
\begin{figure}[!hbt]
\centering{\epsfig{figure=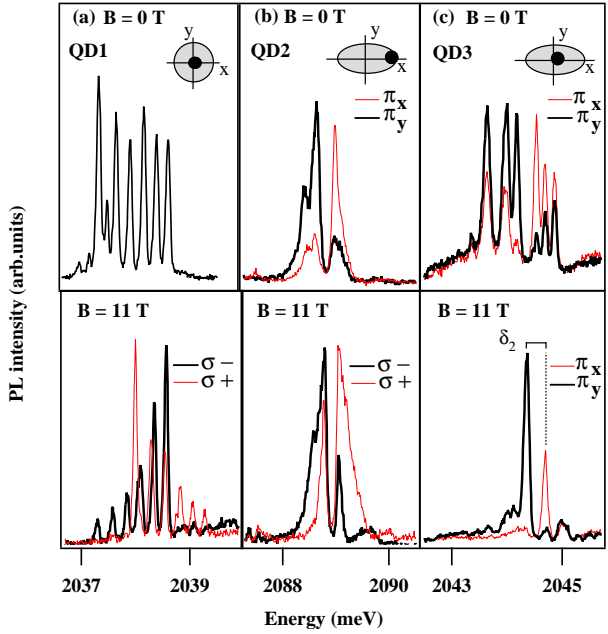,width=8cm}} \caption{(color
online). Low temperature (T=5K) PL spectra of three different
Mn-doped CdTe/ZnTe QDs at B=0T and B=11T. $\sigma_\pm$ are circular
polarizations and $\pi_{x,y}$ are two orthogonal linear
polarizations. (a) Exciton strongly coupled with a Mn atom, in a
symmetric dot. (b) Exciton less strongly coupled with a Mn atom, in
an asymmetric dot. (c) Exciton strongly coupled with a Mn atom, in
an asymmetric dot.}\label{fig1}
\end{figure}

Fig.\ref{fig1} shows the three types of emission spectra obtained at
5K, under low excitation density, for single QDs containing a single
Mn. In the first example (QD1), a structure composed of six main
lines dominates the emission spectrum at zero magnetic field. Such a
spectrum is a fingerprint of a confined exciton interacting with a
single Mn atom \cite{bes04}. These lines correspond to the radiative
(``bright") exciton states $J_{z} = \pm1$ coupled to the six spin
components of the Mn atom ($S=5/2$). Analysis of the line
intensities gives the occupation probability of the six Mn spin
states \cite{bes04}.  The three low intensity lines on the low
energy side of the structure (Fig.\ref{fig1}(a)) can be attributed
to the contribution of exciton dark states \cite{bes04}. We will see
that a fine structure with well separated lines requires not only
that the Mn atom interact strongly with the exciton in the dot, but
also that the dot must retain high symmetry.

By contrast, the emission of QD2 consists of two broad peaks with
width about $200 \mu$eV, separated by an energy gap of about $300
\mu eV$. A similar gap is seen clearly in the third example (QD3),
for which six lines are observed, but with two sets of three lines
separated by the central gap. An additional essential property is
that the emission lines are linearly polarized along two orthogonal
directions for QD2 and QD3, whereas for QD1 the emission is
unpolarized.

Such large differences in the zero field emission spectra can be
attributed to competition between the X-Mn interaction and the
anisotropic e-h exchange interaction arising in asymmetric QDs. We
recall that for non-magnetic QDs, the e-h exchange interaction in an
anisotropic potential mixes the bright exciton states $J_z=\pm1$.
The emission of the QD is then linearly polarized along two
orthogonal directions and split by the anisotropic exchange energy
$\delta_2$ \cite{kul99,ivc97}, originating from the long range
(non-analytic) e-h exchange interaction
$H_{e-h}^{aniso}$\cite{pik71}.

\begin{figure}[!hbt]
\centering{\epsfig{figure=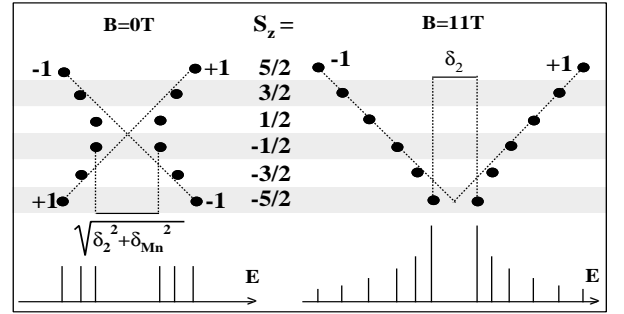,height=4.2cm,width=8cm}}
\caption{Bright state transitions in an asymmetric Mn-doped QD at
B=0T and 11T. The e-h exchange interaction induces anticrossings
(energy $\delta_2$) of the $\pm1$ excitons associated with
$S_z=-1/2, -3/2, -5/2$ for successive values of B; the $S_z=-5/2$
anticrossing is shown at right.}\label{fig2}
\end{figure}

In symmetric Mn-doped QDs, the X-Mn exchange interaction lifts the
degeneracy of the bright exciton states, the splitting being
proportional to the Mn spin projection $S_z$ \cite{bes04}. Such
systems can be described by an effective  Hamiltonian ${H = I_e \;
\emph{\textbf{s}}_e.\textbf{\emph{S}}  + I_h \;
\emph{\textbf{j}}_h.\textbf{\emph{S}}+H_{e-h}^{iso}}$ operating in
the basis of the heavy hole exciton states and Mn spin states. Here,
$I_e$ ($I_h$) is the electron (hole)-Mn exchange integral and
$\emph{\textbf{s}}_e$, $\emph{\textbf{j}}_h$ and $\emph{\textbf{S}}$
are respectively the electron, heavy hole and Mn spins.
$H_{e-h}^{iso}$ is the isotropic part of the e-h exchange
interaction. In zero magnetic field, there are six doubly degenerate
radiative energy levels formed by associating the six Mn spin
projections with the two bright exciton states $J_z = \pm1$ (the
corresponding transitions lie on the dotted cross in
Fig.\ref{fig2}). They are separated by equal energy intervals
$\delta_{Mn} = \frac{1}{2}(I_e-3I_h)$.
We now consider how the simultaneous presence of $H_{e-h}^{aniso}$
and the X-Mn exchange interaction affects the zero field spectrum.
Diagonalizing the augmented Hamiltonian $H+H_{e-h}^{aniso}$ shows
that the e-h exchange interaction splits the six line structure into
two subsets of three lines. Fig.\ref{fig2} represents the bright
state transitions ($J_z=\pm 1$) associated with the six manganese
spin projections $S_z$. The anisotropic e-h exchange term $\delta_2$
mixes the bright exciton states associated with the same Mn spin
projection, inducing an extra splitting between them. The energy
splitting of the bright excitons for a given value of $S_z$ is given
by :
\begin{equation} \label{deltaE}
\begin{array}{cc}
\Delta E (S_z) = \sqrt{{\delta_2}^2+(2\mid S_z\mid  \delta_{Mn})^2}\\
\end{array}
\end{equation}
where $2\mid S_z\mid  \delta_{Mn}$ is the splitting induced by the
Mn only. The mixing induced by $\delta_2$ is thus strongest for the
central pair of states, associated with the $S_{z}=\pm1/2$ Mn spin
projections. Eq.\ref{deltaE} for $\Delta E (S_z)$ shows that
anisotropy of the dot creates a gap in the center of the emission
structure, see Fig.\ref{fig2}. This explains the line patterns
observed for the three types of QD spectra presented above.

The width of the central gap ($\sqrt{{\delta_2}^2+{\delta_{Mn}}^2}$)
and the width of the whole six-line spectrum structure
($\sqrt{\delta_2^2+(5\;\delta_{Mn})^2}$) yield the values of
$\delta_2$ and $\delta_{Mn}$ for the three  dots of Fig.\ref{fig1}.
The ratio $\delta_2/\delta_{Mn}$ determines the spectrum type. For
QD1, the ratio $\delta_2/\delta_{Mn}$ is found to be $<0.2$, with
$\delta_{Mn}=250\mu$eV : the X-Mn interaction dominates the emission
structure and the Mn spin states are directly resolved in the
optical spectrum. By contrast, $\delta_2/\delta_{Mn}= 2.3$ for QD2
($\delta_{Mn}=120\mu$eV), so the anisotropy splitting predominates :
we observe only two broad peaks separated by a central gap and can
no longer resolve the Mn spin states. A reduction in $\delta_{Mn}$
of only a factor of two has completely changed the type of spectrum.
QD3 demonstrates the intermediate case ($\delta_2/\delta_{Mn}=1.3$
with $\delta_{Mn}=230\mu$eV) where the combined effect of the e-h
and X-Mn exchange interactions is seen very clearly : despite the
importance of anisotropy, the Mn spin states are still resolved as
two subsets of three lines separated by the central gap.

The parameter $\delta_{Mn}$ is determined \cite{bhat03,kyr04} by the
values of ${I_e = \alpha\mid\phi_e(R_{Mn})\mid^2}$ and ${I_h =
\beta/3\mid \phi_h(R_{Mn})\mid^2}$. Here $\phi_e$ ($\phi_h$) is the
electron (hole) envelope function, which falls off with $R_{Mn}$,
the distance of the Mn atom from the QD center ; $\alpha$ ($\beta$)
are the Mn-e(h) exchange constants \cite{fur88}. Clearly, for QD2,
the Mn atom is farther from the dot center than for QD1 and QD3.
\begin{figure}[!hbt]
\centering{\epsfig{figure=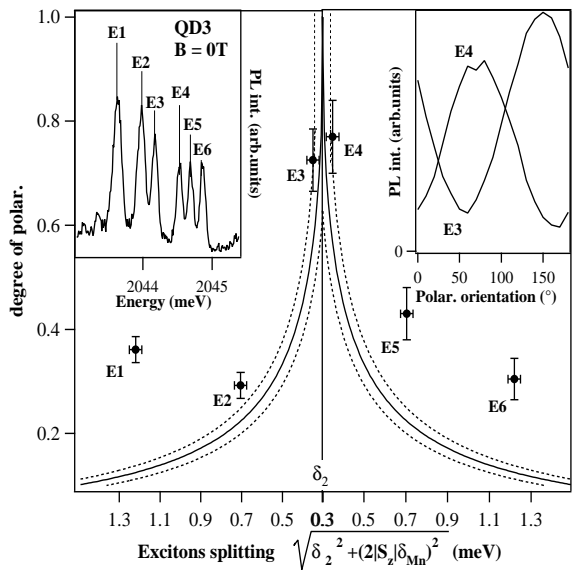,width=7.5cm,height =7.5cm}}
\caption{Measured degree of linear polarization of emission lines
E1-E6 for QD3 in zero field, as a function of the splitting between
the ${\mid J_z=\pm1, S_z>}$ states, compared to theoretical behavior
(full curves ; dotted curves show the uncertainty range
corresponding to the imprecision in $\delta_2$). The left inset
labels the emission lines. The right inset shows the PL intensity of
the  lines E3 and E4 ($S_z=\pm1/2$), as a function of the linear
polarizer orientation. } \label{fig3}
\end{figure}
The theory also explains the linear polarization that we observe in
zero magnetic field. When anisotropic electron-hole exchange
interaction is included, the eigenstates of the X-Mn system are of
the form \cite{ivc97,bes00} :
\begin{equation}\label{vecp}
\begin{array}{cc}
\mid+>\mid S_z>=(cos \theta \mid+1>+sin \theta\mid-1>)\mid S_z>,\\
\mid->\mid S_z>=(cos \theta\mid-1>-sin \theta\mid+1>)\mid S_z>,
\end{array}
\end{equation}
where $tan 2\theta = \delta_2/(2\mid S_z\mid\delta_{Mn})$. The
mixing of the bright states associated with a Mn spin state $S_z$ is
controlled by the ratio of $\delta_2$ to the Mn induced splitting
$2\mid S_z\mid \delta_{Mn}$. The emission lines have partial linear
polarization along two orthogonal directions corresponding to the
principal axes of the anisotropic potential.  Fig.\ref{fig3}
compares measured and theoretical degree of polarization for QD3.
The data are in qualitative agreement with the theoretical curve. In
particular, central lines E3 and E4, associated with $S_z=\pm1/2$,
are almost completely polarized. The expected decrease with increase
of $\Delta E (S_z)$ is reproduced qualitatively especially for the
three upper energy lines E4, E5 and E6. For the lower energy lines,
the degree of polarization could be influenced by the non radiative
states ($J_{z} =\pm2$) which lie in their energy range.

\begin{figure}[!hbt]
\centering{\epsfig{figure=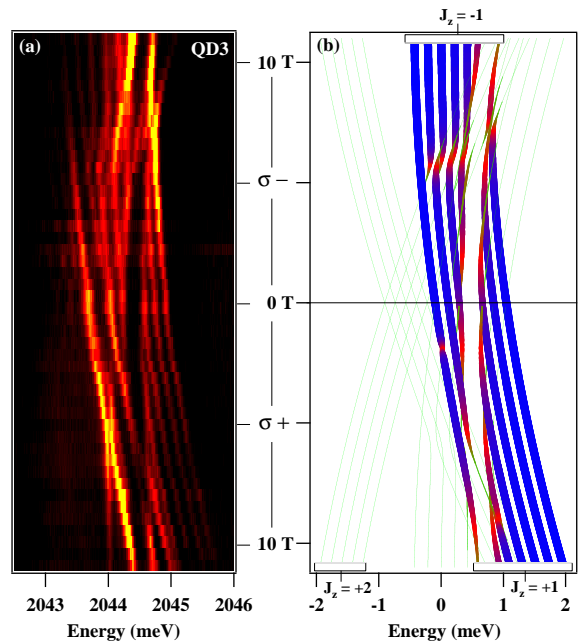,width=7.6cm}} \caption{(color
online). (a) Intensity map of magnetic field dependence of the
emission spectrum of asymmetric Mn-doped dot QD3, for circular
polarization $\sigma+$ and $\sigma-$. (b) Optical transitions
obtained from the diagonalization of the
spin$+$Zeeman$+$diamagnetism Hamiltonian in the subspace of the 24
heavy-hole exciton and Mn spin components; Line thickness and color
scale for $\sigma+$, $\sigma-$ are proportional to absolute value of
the projection of the exciton state on the $J_z = +1,-1$ exciton
respectively(green$=$low intensity, blue$=$high intensity). The two
transitions which are forbidden at all magnetic fields (${\mid
J_z=\pm2,S_z=\mp5/2>}$) are not plotted.} \label{fig4}
\end{figure}
\begin{figure}[!hbt]
\centering{\epsfig{figure=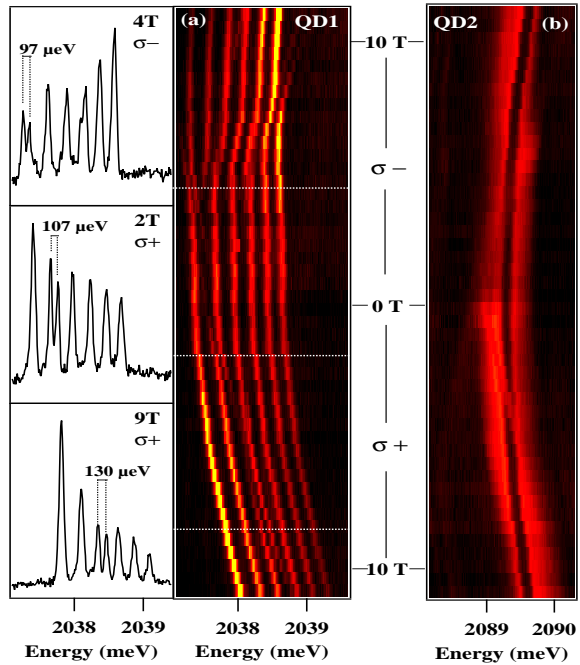,width=7.6cm,height =8.8cm}}
\caption{(color online). Magnetic field dependence of the emission
of QD1 and QD2. (a)Exciton strongly coupled with a single manganese
atom, in a quasi-symmetric dot . Anticrossings discussed in the text
are illustrated by the spectra in the left insets. (b)Exciton less
strongly coupled with a manganese atom, in a symmetric dot .}
\label{fig5}
\end{figure}

Our interpretations of the above spectra are confirmed by
magneto-optical measurements (Fig. \ref{fig1}, \ref{fig4} and
\ref{fig5}). For QD3 (Fig. \ref{fig4}(a)) and QD1 (Fig.
\ref{fig5}(a)), the typical Zeeman splitting of the six lines is
clearly observed in the data at all fields, with a strong intensity
gradient at the highest fields (see Fig.\ref{fig1}) resulting from a
rather strong Mn spin polarization. For the clearly anisotropic dots
(QD3 and QD2, Fig.\ref{fig5}(b)), the central gap in the emission
structure is maintained in both circular polarizations, with a small
quadratic diamagnetic energy shift. This behavior is explained as
follow. The dot anisotropy leads to successive anticrossings of the
$\pm1$ exciton states associated with given Mn spin projections
($-1/2$, $-3/2$ and $-5/2$) as a function of magnetic field : As B
increases, transitions associated with the $J_z=+1$ exciton shift up
in energy whereas the $J_z=-1$ transitions shift down (see
Fig.\ref{fig2}). The anisotropic part of the electron-hole exchange
interaction mixes successively the $J_z=\pm1$ exciton states
associated with $S_z=-1/2$, then with $S_z=-3/2$ and finally with
$S_z=-5/2$ at successively higher B. For QD3, these anticrossings
are observed successively at 2.5, 7 and 11T.
To understand fully the rich magnetic behavior of dots like QD3, we
calculated the optical transitions under magnetic field by
diagonalizing the complete Hamiltonian of the electron-heavy hole-Mn
system (including the exchange, Zeeman and diamagnetism
Hamiltonians). Calculated transitions are presented in
Fig.\ref{fig4}(b). The fitted Land\'e factors of the electron ($g_e
= -1.1$), the hole ($g_h = 0.3$) and the Mn atom ($g_{Mn}= 2.0$),
the splitting between $J_z=\pm1$ and $J_z=\pm2$ excitons ($ = 1$meV)
and the diamagnetic factor ($\gamma = 2.45 \mu$eV.T$^{-2}$) agree
well with previous work \cite{bes04,fur88,bes00}. Parameters
$\delta_2$ and $\delta_{Mn}$ were adjusted to fit the zero field
data, as explained earlier.

Comparison between calculation and data explains most of the details
of the magneto-optic properties of QD3. In particular, around 7T,
the central gap is perturbed in both circular polarizations. In
$\sigma-$, this is due to anticrossings induced by the
 mixing  of ${\mid s_{ez} =1/2,j_{hz}=-3/2,S_z>}$
states and ${\mid-1/2,-3/2,S_z+1>}$ states by the
\textit{electron}-Mn exchange, \cite{bes04}, i.e. corresponding to
simultaneous spin-flips of electron and manganese spins. In
$\sigma+$ polarization, Fig.\ref{fig4}(b) shows that the line of
second lowest energy crosses the central gap as an essentially
non-radiative transition. This implies a mixing of
${\mid-1/2,3/2,-3/2>}$ and ${\mid-1/2,-3/2,-1/2>}$. This is a second
order mixing involving both  mixing of ${\mid-1/2,-3/2,-1/2>}$ and
${\mid1/2,-3/2,-3/2>}$ by the e-Mn exchange and mixing of
${\mid-1/2,3/2,-3/2>}$ and ${\mid1/2,-3/2,-3/2>}$ by the anisotropic
e-h exchange ; that is, the e-Mn exchange induces a mixing of states
mediated by the anisotropy-induced coupling.

We now consider the field dependence of the two extreme cases
illustrated by QD1 and QD2. For QD2 (Fig.\ref{fig5}(b)), where
$\delta_2/\delta_{Mn}$=2.3, the field dependence is mainly the
quadratic variation of the central gap. At $B=11$T, the Zeeman
splitting has nevertheless separated the $\pm1$ excitons and the
circular polarization is almost completely restored
(Fig.\ref{fig1}(b)). The perturbation of the central gap around 4T
in $\sigma-$ polarization can be attributed, as for QD3, to the
mediating electron-hole exchange.

QD1 presents weak anticrossings (see insets in Fig.\ref{fig5}(a))
that are not explained by the simple model of a Mn-doped symmetric
QD. Comparing with calculations we can attribute these anticrossings
to a slight anisotropy, too small to be revealed by the presence of
a clear central gap at zero field: the perturbation occurring from 5
to 9T in $\sigma+$ polarization corresponds to a mixing between
bright states associated with Mn spin projection $-1/2$, and
represents a residue of the central gap discussed extensively above.
The other anticrossings involve bright ($J_{z} =\pm1$) and dark
($J_{z} =\pm2$) states  and are due to second order electron-Mn
exchange interaction mediated by the anisotropic e-h exchange
interaction, as seen more clearly in the strongly anisotropic QDs.

In summary, we demonstrate that the position of a single Mn atom in
a QD is not the only parameter that has to be controlled in order to
resolve the Mn spin states. Another geometric parameter must be
considered: the asymmetry of the dot. The interplay between these
two parameters has important consequences for the QD optical
properties. The Mn-exciton exchange interaction tends to separate
the  bright exciton states, whereas the anisotropic part of the
electron-hole exchange interaction tends to couple them and to hide
the Mn spin splitting. The wealth of data obtained gives a unified
picture of the effects of dot asymmetry on the fine structure and
polarization properties of optical transitions in single Mn-doped
QDs. This allows us to determine the conditions required to tune the
magnetic QD states in order to control and manipulate individual
localized spins by single carriers.

The authors are grateful to J.Cibert and R. Cox for fruitful
discussions and to Y. Genuist and M. Falco for expert technical
assistance.


\end{document}